\documentclass[prd%
,aps%
,twocolumn%
,amssymb%
,showpacs%
,preprintnumbers%
,nofootinbib%
]{revtex4}
\usepackage{epsfig}
\begin{document}

\def\lsi{\raise0.3ex\hbox{$<$\kern-0.75em\raise-1.1ex\hbox{$\sim$}}}
\def\gsi{\raise0.3ex\hbox{$>$\kern-0.75em\raise-1.1ex\hbox{$\sim$}}}
\newcommand{\lsim}{\mathop{\lsi}}
\newcommand{\gsim}{\mathop{\gsi}}
\def\hxi{\hat{\xi}}
\def\hk{\hat{k}}
\def\bE{\mbox{\boldmath ${E}$}}
\def\bB{\mbox{\boldmath ${B}$}}
\def\bD{\mbox{\boldmath ${D}$}}
\def\vA{\vec A}
\def\vE{\vec E}
\def\vB{\vec B}
\def\vD{\vec D}
\def\cH{{\mathcal H}}
\def\half{\frac{1}{2}}
\def\archive#1{ #1}

\title{Magnetic monopoles from gauge theory phase transitions}

\author{A.~Rajantie}
\affiliation{DAMTP, CMS, University of Cambridge, Wilberforce Road,
Cambridge CB3 0WA, United Kingdom}

\date{11 July, 2003}

\begin{abstract}
Thermal fluctuations of the gauge field lead to monopole formation 
at the grand unified phase transition in the early universe, even if
the transition is merely a smooth crossover. 
The dependence of the produced monopole density on various parameters 
is qualitatively different from theories
with global symmetries, and the monopoles have a positive correlation
at short distances. 
The number density of monopoles may be suppressed if the
grand unified symmetry is only restored for a short time by,
for instance, non-thermal symmetry restoration after preheating.

\end{abstract}
\pacs{PACS: 11.15.Ex, 11.27.+d, 74.60.Ge}

\preprint{DAMTP-2002-157}
\preprint{hep-ph/0212130}
\maketitle

It is a generic property of grand unified theories (GUTs) that
magnetic monopoles of mass of the order $m_M\approx 10^{16}$~GeV 
exist~\cite{'tHooft:1974qc,Polyakov:ek}, and these
monopoles would have been produced in large 
numbers in the GUT phase transition at 
$T_{\rm GUT}\approx m_M$~\cite{Kibble:1976sj}.
Afterwards, pair annihilations can decrease the monopole density,
but
estimates show that the number density would still be
comparable to baryons~\cite{Preskill:1979zi}. 
Because the monopoles are $10^{16}$ times heavier than protons,
this would have caused the universe to collapse under its own weight
long ago.

This 
monopole 
problem, alongside with several other cosmological puzzles, was
wiped away by the theory of inflation~\cite{Guth:1980zm},
as the monopole density would have been diluted to a negligible
level by
a period of accelerating expansion.
For this to solve the problem, the reheat temperature at which the
universe thermalizes
must be lower than $T_{\rm GUT}$. 
These constraints are even stronger
in models with non-perturbative effects such as
preheating~\cite{Kofman:1995fi}, since the GUT symmetry can be
temporarily
restored \cite{Khlebnikov:1998sz,Rajantie:2000fd} 
and topological defects formed
even if the reheat temperature is well below 
$T_{\rm GUT}$~\cite{Kasuya:1998td,Tkachev:1998dc}.
It is therefore important to understand how monopoles are formed
to estimate how strong the
bounds imposed by the monopole problem really are.

In this paper, I will discuss monopole formation at a phase transition
that starts from complete thermal equilibrium.
It is clear that this is not actually the case 
for the GUT transition, because of the
high expansion rate and the non-equilibrium effects mentioned above.
Nevertheless, the assumption of thermal equilibrium simplifies the
problem significantly and makes it possible to identify the physical
mechanisms that are responsible for monopole formation. Once these
mechanisms are understood, their effects can be studied in more
realistic non-equilibrium settings.

The symmetry broken at the GUT phase transition is a local gauge
invariance, whereas
most of the existing
literature on monopole
formation 
implicitly assumes a breakdown of a global 
symmetry. The Kibble (or Kibble-Zurek)
mechanism~\cite{Kibble:1976sj,Zurek:1985qw}, 
which forms the monopoles in the global case,
is ultimately based on the observation that the direction of the 
order parameter cannot be
correlated at infinitely long distances.
Because the direction of the order
parameter is not gauge invariant,
this argument cannot be used in GUTs.

Moreover, gauge symmetries cannot be spontaneously broken~\cite{Elitzur:im}.
For rather 
generic parameter values, there is actually no phase transition at
all, 
but simply a
smooth crossover between the phases~\cite{Polyakov:vu,Hart:1996ac}. 
Does this mean that the whole evolution could be
adiabatic and thereby there would
be no monopole formation at all?

I will present an argument
that shows that monopoles are still formed. 
This result is based on causality and 
the conservation of magnetic charge. In
fact, Weinberg and Lee~\cite{Weinberg:uq,Lee:gt} have used somewhat
similar reasoning to constrain later annihilations of monopoles after
the phase transition in the context of the Kibble mechanism.
As I will argue, there are long-wavelength thermal fluctuations of the
magnetic charge in the symmetric phase, and they will freeze out
forming monopoles. These fluctuations are physical and well defined,
because the high and low
temperature phases are smoothly connected.
As there is, in this sense, 
a high density of monopoles above the transition, one
could say that we are describing annihilation rather than formation of
monopoles.
This is obviously a matter of taste, but in any case,
the monopoles do not correspond to localized energy
concentrations in the symmetric phase and cannot therefore be thought
of as particles.

The mechanism presented in this paper is physically different
from global theories. Both involve a freeze-out of
long-wavelength
degrees of freedom, but in global theories this happens
when the scalar correlation length
diverges at the transition point.
In our case, everything is finite at the (approximate) transition point but
the magnetic screening length 
diverges in the
zero-temperature limit. The reason why this freeze-out leads to
monopole formation is also different.
As we shall see, 
the monopoles formed in a gauge theory
have positive correlations at short distances,
which is the opposite of what the Kibble mechanism 
predicts~\cite{Digal:1998ak}.
The number density of monopoles will also be
qualitatively different from the Kibble mechanism.


Let us start by briefly reviewing the standard Kibble mechanism
\cite{Kibble:1976sj}%
, which is valid in the case of global symmetries%
.
For simplicity, we shall discuss the SU(2) symmetry group only, but
the same arguments should apply to SU(5), SO(10) or other possible
GUTs.
The Lagrangian of the 
global SU(2) scalar
theory is
\begin{equation}
{\cal L}={\rm Tr}\partial_\mu\Phi\partial^\mu\Phi
-m^2{\rm Tr}\Phi^2-\lambda\left({\rm Tr}\Phi^2\right)^2,
\label{equ:global}
\end{equation}
where $\Phi$ is in the adjoint representation, and we are
assuming that at zero temperature, the SU(2) symmetry is broken. To
leading order, this means $m^2<0$.

We shall consider this theory at a non-zero temperature $T$. When the
temperature is high enough, the SU(2) symmetry is unbroken. We are asking
what happens if we start from thermal equilibrium in the
symmetric phase and gradually decrease the temperature so that the
symmetry gets broken.
As long as couplings are weak, we can approximate the equilibrium and
near-equilibrium dynamics reasonably well
by a classical field theory with a
temperature-dependent mass term 
$m^2(T)$~\cite{Farakos:1994kx,Moore:1996bn,Rajantie:1997pr,Hindmarsh:2001vp}.
Although it is difficult to make this approach quantitatively very
accurate, it gives us a way of thinking about the dynamics in terms of 
hot classical fields without quantum mechanical complications.

It only takes a small change in temperature
near the critical temperature $T_c$ to cause the phase transition, and
this effect is mainly due to the effective mass parameter's changing
from positive to negative. Therefore, we shall
simply consider keeping $T$ fixed and varying $m^2$.

Let us now discuss the dynamics of the global theory
(\ref{equ:global}).
In the high-temperature phase, the
field $\Phi$ vanishes on the average. In the broken phase, it would
ideally have a non-zero constant value $\Phi(\vec{x})=\Phi_0$, where
${\rm Tr}\Phi_0^2=\phi^2/2=-m^2/2\lambda>0$, but
this would require ordering of the
field at infinite distances, which cannot be achieved in finite
time. Instead, the scalar correlation length $\xi$ grows as the transition
point is approached, but freezes out to some finite value
$\hat{\xi}$, which
is determined by
the critical dynamics of the
system~\cite{Zurek:1985qw} and ultimately limited by causality.

After the transition, 
we can imagine that the system consists of domains of radius
$\hat\xi$,
between which
$\Phi$ is totally uncorrelated. At each point where four of these domains
meet, there is a fixed, non-zero probability that the field cannot
smoothly interpolate between the domains without vanishing at a
point. This point is a monopole, and therefore this scenario predicts
a monopole density 
\begin{equation}
n_M^{\rm Kibble}\approx \hat\xi^{-3}.
\label{equ:kibblepred}
\end{equation}
Furthermore, there is a strong negative
correlation between monopoles at short 
distances~\cite{Digal:1998ak,Rajantie:2001ps}: 
Imagine a sphere
centred at a monopole. If the radius is greater than $\hat\xi$, each
point at the sphere is uncorrelated with its centre and therefore
insensitive to whether there is a monopole inside or not. Consequently,
the average winding number must be zero and there
must be an antimonopole within distance $\approx\hat\xi$ from each monopole.

Having reviewed the Kibble-Zurek scenario,
let us now turn our attention to the gauge theory. 
The gradients in the Lagrangian are replaced by covariant derivatives
$D_\mu=\partial_\mu+igA_\mu$,
\begin{eqnarray}
{\cal L}&=&
-\frac{1}{2}{\rm Tr}F_{\mu\nu}F^{\mu\nu}+
{\rm Tr}[D_\mu,\Phi][D^\mu,\Phi]
\nonumber\\&&
-m^2{\rm Tr}\Phi^2-\lambda\left({\rm Tr}\Phi^2\right)^2,
\end{eqnarray}
where $F_{\mu\nu}=(ig)^{-1}[D_\mu,D_\nu]$.

At a high temperature $T$ and at weak coupling $g$,
the phase structure of this theory is given to a good approximation
by a three-dimensional effective theory~\cite{Rajantie:1997pr}, 
%
which depends on two
parameters, $g^{-2}(T/T_c-1)$ and
the ratio of the coupling constants $\lambda/g^2$.
There is a line of first-order transitions at
small $\lambda/g^2$~\cite{Coleman:jx}, which ends at a
second-order point at around 
$\lambda/g^2\approx 0.3$~\cite{Kajantie:1997tt}.
To simplify the estimates, we actually assume that
$\lambda/g^2\approx 1$, which is above the critical value so that
the two phases are smoothly connected~\cite{Hart:1996ac} and
all correlation lengths are finite.
In particular, the scalar correlation length is 
microscopic~\cite{Kajantie:1997tt}, 
of order $1/g^2T$, and therefore we ignore scalar fluctuations.
Even then, it is still possible to find an approximate
crossover point, 
which separates a ``symmetric'' and a ``broken''
phase, and we will use this terminology although it is not quite
precise.

In the broken phase,
there is still an unbroken U(1) subgroup, and the corresponding
magnetic field is given by the 't~Hooft operator~\cite{'tHooft:1974qc},
\begin{equation}
{\cal B}_i=\frac{1}{2}\epsilon_{ijk}
\left[
{\rm Tr}\hat\Phi F_{jk} + \frac{1}{2ig} {\rm Tr}
\hat\Phi (D_j\hat\Phi) (D_k\hat\Phi) \right],
\end{equation}
where $\hat\Phi=\Phi\sqrt{2/{\rm Tr}\Phi^2}$ is well-defined almost 
everywhere.
In continuum, ${\cal B}_i$ is sourceless apart from points where
$\Phi$ vanishes, and has sources of integer multiples of $4\pi/g$ at
those points. Being sources of the magnetic field, these points
are quantized magnetic charges, and therefore we call them magnetic
monopoles, whether or not they resemble the 't~Hooft-Polyakov monopole
solution~\cite{'tHooft:1974qc,Polyakov:ek}. It is straightforward to
see that magnetic charge defined in this way is conserved, i.e., the
world lines of these monopoles cannot end.

The same construction has also been done on
lattice~\cite{Kronfeld:1987vd,Davis:2000kv}, and the conservation law
of magnetic charge is preserved. Even
though we will not discuss numerical simulations in this paper, 
this is very important, because classical field theory cannot be in
thermal equilibrium in continuum. Because the monopoles are
well-defined and stable on lattice, we can consistently talk about
them and the magnetic field at a non-zero temperature.

Deep in the broken phase, where
$m_M> T$, we can treat the monopoles as point-like
particles. Therefore, we have the standard expression for the equilibrium
monopole density
\begin{equation}
n_M^{\rm eq}\approx
(m_MT)^{3/2}\exp\left(-\frac{m_M}{T}\right).
\label{equ:eqdens}
\end{equation}
The monopole mass $m_M$ is roughly $m_M\approx \phi/g \approx
(-m^2/\lambda g^2)^{1/2}$.
When $m^2$ decreases further, $m_M$ grows
rapidly, which suppresses $n_M^{\rm eq}$.

If the monopoles did not have long-range interactions, they
would be essentially uncorrelated and behave very
much like the magnetic field in the Abelian Higgs 
model~\cite{Hindmarsh:2000kd,Rajantie:2001ps}, 
albeit in three rather than two
dimensions. 
There is, however, a magnetic Coulomb interaction between the
monopoles, and we shall see that it suppresses their production.
This interaction gives rise to correlations, which are
reflected in the screening of the magnetic field by the 
monopoles~\cite{Polyakov:vu,Davis:2001mg}, in
analogy with the Debye screening of the electric field.

We define the
magnetic screening length
$\xi_B$ as the decay rate of the correlator of 't~Hooft
field strength operators ${\cal B}_i$. 
In equilibrium, 
it is approximately
\begin{equation}
\xi_B\equiv 1/m_B\approx\sqrt{\frac{T}{n_Mq_M^2}}
\approx\sqrt{\frac{g^2T}{n_M}},
\label{equ:debye}
\end{equation}
where $q_M=4\pi/g$ is the magnetic charge of a monopole.
Correspondingly, 
if we define
\begin{equation}
\rho_M=\vec{\nabla}\cdot\vec{\cal B},
\end{equation}
the magnetic charge-charge correlator is
\begin{equation}
\langle\rho_M(\vec{x})\rho_M(\vec{y})\rangle
\approx q_M^2n_M\left(\!
\delta(\vec{x}\!-\!\vec{y})-\frac{m_B^2}{4\pi|\vec{x}\!-\!\vec{y}|}
e^{-m_B|\vec{x}\!-\!\vec{y}|} \right)\!.
\label{equ:eqcorrcoord}
\end{equation}
Using Eq.~(\ref{equ:eqdens}), we find that the
equilibrium screening length behaves as
\begin{equation}
\xi_B\approx \left(\frac{g^4}{Tm_M^3}\right)^{1/4}e^{m_M/2T}.
\label{equ:eqxiB}
\end{equation}
Because there is no phase transition, the correlators of ${\cal B}_i$
or $\rho_M$ cannot change qualitatively when we move to the symmetric
phase. Otherwise, their behaviour could be used to distinguish
between the phases. Consequently, the screening length is always well 
defined, and we can
actually use Eq.~(\ref{equ:debye}) to define the monopole density $n_M$
in the symmetric phase.
Furthermore, we expect that
above the crossover, $\xi_B\approx (g^2T)^{-1}$, because 
the only relevant scale for equal-time correlations 
is $g^2T$~\cite{Hart:1996ac}.\footnote{%
The scale $g^2T$ is generic in hot gauge theories and is
known as the magnetic screening scale. This magnetic screening refers
to non-Abelian magnetic fields, and it is uncertain whether it can be
thought of originating from some kind of monopoles. In contrast, the
field ${\cal B}_i$ is, by definition, screened by monopoles, because
the monopoles were defined as sources of ${\cal B}_i$.}

If $m^2$ is decreased at a constant rate, $\xi_B$ would have to grow
exponentially fast to stay in equilibrium,
but, obviously,
it cannot grow faster than the speed of light. In practice, 
it would grow much slower than this.
This means that sooner or later the growth rate 
$d\xi_B/dt$ needed for
the system to stay in equilibrium exceeds the maximum value, and the
system falls out of equilibrium. We shall denote the time when
this happens by $\hat{t}$. 

The screening length $\xi_B$ can
still keep on growing, but so slowly that we can ignore it if we are
only interested in finding an order-of-magnitude estimate for the initial
monopole density. Therefore, we define the freeze-out screening
length $\hat\xi_B$ as $\xi_B$ at the time when it falls out of
equilibrium,
\begin{equation}
\hat\xi_B=\xi_B(\hat{t}).
\end{equation}
At the time of the freeze-out, the monopole density is
\begin{equation}
\hat{n}_M\approx \frac{T}{q_M^2\hat{\xi}_B^2}
\approx \frac{g^2T }{ \hat{\xi}_B^2}.
\label{equ:freezeoutdensity}
\end{equation}
Note that by cooling the system slower, we can make $\hat\xi_B$
arbitrarily large.
When $\hat{\xi}_B\gg (g^2T)^{-1}$, 
the typical distance 
$\hat{d}\approx \hat{n}_M^{-1/3}$ between monopoles
and antimonopoles is much shorter than the screening length.

Even after the freeze-out, the monopole density will keep on
decreasing, but this is now due to pair annihilations at length scales
shorter than $\hat\xi_B$.
These annihilations smoothen the distribution of monopoles at short 
distances, but they cannot remove them completely~\cite{Weinberg:uq,Lee:gt}. 
To see this, consider a sphere of radius $\hat\xi_B$.
The annihilations may reduce the number of monopoles inside the sphere 
to the
minimum, but they cannot change
its net magnetic charge significantly. 
While the net magnetic charge is zero on the average, it fluctuates
with a root-mean-squared value of
\begin{equation}
Q_M(\hat\xi_B)
= \sqrt{\left\langle \left(\int^{\hat\xi_B} d^3x \rho_M(\vec{x})
\right)^2 \right\rangle}\approx \sqrt{T\hat\xi_B}.
\label{equ:rmscharge}
\end{equation}
Since the annihilations cannot
reduce the charge below this, the monopole
density cannot fall below
\begin{equation}
n_M\approx \frac{Q_M(\hat\xi_B)}{q_M\hat\xi_B^3} 
\approx q_M^{-1}\sqrt{\frac{T}{\hat\xi_B^5}}\approx 
g\sqrt{\frac{T}{\hat\xi_B^5}}.
\label{equ:omapred}
\end{equation}
We have not shown how to estimate $\hat\xi_B$, but
nevertheless, this expression is clearly different from the
Kibble-Zurek result (\ref{equ:kibblepred}), 
because of the explicit appearance of
$g$ and $T$.

Moreover, as long as $Q_M(\hat\xi_B)\gg q_M$, there
will be clusters of monopoles of equal sign, and the number of
monopoles in each of them can be large if $T\gg \hat\xi_B^{-1}$.
This means that there is a positive correlation between monopoles
at short distances, very much in the same way as in the case of
vortices in the Abelian Higgs model~\cite{Hindmarsh:2000kd,Rajantie:2001ps}
and in stark contrast with
the Kibble mechanism. 

We can reach the same conclusions by studying the time evolution of
the magnetic charge correlator in the Fourier space. We define 
the equal-time correlator
$G(k)$
by
\begin{equation}
\langle \rho_M(\vec{k})\rho_M(\vec{q})\rangle
=q_M^2 G(k)(2\pi)^3\delta\left(\vec{k}+\vec{q}\right),
\end{equation}
and from Eq.~(\ref{equ:eqcorrcoord}), we find
\begin{equation}
G(k)=\frac{T}{q_M^2}\frac{m_B^2k^2}{k^2+m_B^2}.
\label{equ:eqG}
\end{equation}
As there is no transition, we expect that
\begin{equation}
G(k)\approx \frac{Tk^2}{q_M^2}
\label{equ:eqGlowk}
\end{equation}
in the symmetric phase where $m_B$ is large.  

Deep in the broken phase, $G(k)$ approaches zero, but
causality implies that very long-wavelength (low $k$) 
correlations can only change
slowly~\cite{Weinberg:uq,Lee:gt}.  
We can give a rough upper bound for the rate of change,
\begin{equation}
\left|\frac{d \ln G(k)}{dt}\right| \lsim k.
\label{equ:adiabgen}
\end{equation}
Using Eq.~(\ref{equ:eqG}), this becomes
\begin{equation}
\frac{k^2}{k^2+m_B^2}\frac{d\ln m_B^2}{dt}\lsim k.
\label{equ:adiabcond}
\end{equation}
Below the transition, $\ln m_B\approx \sqrt{-m^2}/g^2T$, and if we keep on
decreasing $m^2$, then sooner or later Eq.~(\ref{equ:adiabcond}) 
ceases to be satisfied
for $k$ less than some critical value $\hat{k}$. The modes
with higher $k$ keep on decreasing and we approximate the final
correlator by
\begin{equation}
G(k)\approx \frac{Tk^2}{q_M^2}\exp\left(-\frac{k^2}{2\hat{k}^2}\right).
\label{equ:freezeoutG}
\end{equation}
A Gaussian fall-off like this would follow naturally from diffusion, but
our conclusions do not depend on the precise form of the correlator,
as long as it has a relatively sharp cutoff at $\hat{k}$.
The corresponding monopole density is given by
\begin{equation}
n_M\approx \left(\int \frac{d^3k}{(2\pi)^3}G(k)\right)^{1/2}
\approx q_M^{-1}\sqrt{T\hat{k}^5},
\end{equation}
which agrees with Eq.~(\ref{equ:omapred}) if we identify $\hat{k}=1/\hat\xi_B$.

We can also find the monopole-monopole correlator in coordinate space
by taking the Fourier transform of Eq.~(\ref{equ:freezeoutG}),
\begin{equation}
G(r)\approx\frac{T\hat{k}^5}{q_M^2}\frac{e^{-r^2\hat{k}^2/2}}{(2\pi)^{3/2}}
\left(
3-r^2\hat{k}^2\right),
\end{equation}
and it is indeed positive at distances
$r\lsim \sqrt{3}/\hat{k}$.

As a concrete example,
let us now estimate the
monopole density produced in the GUT phase transition
using only causality to limit the growth of $\xi_B$.
It is clear that causality leads to a freeze-out, because the currect
magnetic screening length would be proportional to $\exp(m_M/2T)\sim
\exp(10^{28})$ and therefore enormously longer than the size of the
observable universe.
This is still an oversimplification and the estimates should not be
taken literally.

At high temperatures, the effective mass parameter of the theory is 
$m^2(T)\approx g^2(T^2-T_{\rm GUT}^2)$.
Because of the expansion of the universe, the temperature is
decreasing at the rate $dT/dt\approx -T^3/M_P$, where $M_P\approx
10^{19}~{\rm GeV}$ is the Planck mass. Near $T_{\rm GUT}$,
we can therefore approximate
\begin{equation}
m^2\approx -g^2\frac{T_{\rm GUT}^4}{M_P}t.
\end{equation}
Deep enough in the broken phase, the monopole mass grows as
\begin{equation}
m_M\approx \sqrt{\frac{t}{g^2M_P}}T_{\rm GUT}^2.
\end{equation}
From Eq.~(\ref{equ:eqxiB}) we see that the growth rate of $\xi_B$ 
is
\begin{equation}
\frac{d\xi_B}{dt}\approx \frac{T_{\rm GUT}^{11/4}}{gm_M^{7/4}M_P}
e^{m_M/2T_{\rm GUT}}=
\frac{T_{\rm GUT}}{gM_P}x^{-7/4}e^x,
\end{equation}
where we have introduced the dimensionless variable $x=m_M/2T_{\rm GUT}$.
We require that this is equal to $1$ for the freeze-out scale, 
and find
$x\approx \ln(gM_P/{T_{\rm GUT}}),
$
and consequently
$
\hat{\xi}_B\approx g^2 M_P/T_{\rm GUT}^2.
$
Then, Eq.~(\ref{equ:omapred})
tells us that the monopole density is
\begin{equation}
n_M\approx \frac{1}{g^4}\left(\frac{T_{\rm GUT}^{11}}{M_P^5}\right)^{1/2},
\label{equ:GUTpred}
\end{equation}
which we can compare with the prediction of the Kibble mechanism under
the same circumstances~\cite{Einhorn:ym},
\begin{equation}
n^{\rm Kibble}_M\approx \frac{g^2T_{\rm GUT}^4}{M_P}.
\end{equation}
The two results differ
by a factor of $g^6(M_P/T_{\rm GUT})^{3/2}$, which is not particularly
large for realistic GUTs, but could in principle have any value.

According to Eq.~(\ref{equ:rmscharge}),
the typical number of monopoles in a cluster is
\begin{equation}
N_M^{\rm net}=\frac{Q_M}{q_M}\approx g^2\sqrt{\frac{M_P}{T_{\rm
GUT}}}.
\label{equ:clustpred}
\end{equation}
This combination is, again, of order one, which means that there is a
possibility of forming small clusters.

As was already mentioned, the estimate in
Eq.~(\ref{equ:GUTpred}) 
is not very precise. The main factor in this is that the magnetic
charges are likely to move diffusively rather than at the speed of
light.
The true freeze-out scale $\hat\xi_B$ is necessarily shorter than our
estimate and therefore Eq.~(\ref{equ:GUTpred}) can be thought of as an
approximate lower bound and Eq.~(\ref{equ:clustpred}) as an upper bound.
Furthermore,
if the transition is fast enough, which 
may actually be the case in the GUT transition, the
approximation in Eq.~(\ref{equ:eqdens}) 
that the monopoles are point particles is
not justified and one should instead use a field theory description.
Nevertheless, 
this simplified calculation shows the places where more accurate physical
input is needed to improve the estimates.

It is also interesting to apply this same picture to 
cases where the GUT symmetry is restored only briefly
after inflation, either because of ``non-thermal''
fluctuations~\cite{Khlebnikov:1998sz,Rajantie:2000fd,%
Tkachev:1998dc,Kasuya:1998td} or because the reheat temperature is
slightly above $T_{\rm GUT}$.
The estimated monopole density depends on the low-momentum behaviour
of $G(k)$ given in Eq.~(\ref{equ:eqGlowk}). Because of charge
conservation, the monopoles and antimonopoles must be produced in
pairs, and even if they move at the speed of light, the leading term
in $G(k)$ grows as $G(k)\sim n_Mk^2t^2$. It will therefore take at
least the time $t_{\rm eq}\approx(g^2T/n_M)^{1/2}\approx 
\xi_B$ to achieve the form
(\ref{equ:eqGlowk}). This conclusion can also be reached by
considering the time it takes for the pairs to reach the 
equilibrium size $\sim\xi_B$.

This means that if the GUT symmetry is restored only very briefly, for
a period shorter than $t_{\rm eq}\approx (g^2T_{\rm GUT})^{-1}$, the number
density of monopoles will be suppressed. In reality, the equilibration
process is probably significantly slower, and therefore $t_{\rm eq}$
can be much larger, perhaps even so large that the bounds on the
reheat temperature disappear completely. 
In any case, a more careful analysis of the
dynamics is needed to estimate how strong the suppression is in
practice and whether it solves the monopole problem in the case of
non-thermal symmetry restoration.

The author would like to thank
Mark Hindmarsh, Tom Kibble and Andrei Linde
for useful discussions, and PPARC, 
Churchill College and the ESF Programme ``Cosmology in the
Laboratory'' for financial support.



\begin{thebibliography}{99}
\bibitem{'tHooft:1974qc}
G.~'t Hooft,
Nucl.\ Phys.\ B {\bf 79}, 276 (1974).

\bibitem{Polyakov:ek}
A.~M.~Polyakov,
JETP Lett.\  {\bf 20}, 194 (1974).

\bibitem{Kibble:1976sj}
T.~W.~B.~Kibble,
J.\ Phys.\ {\bf A9}, 1387 (1976).

\bibitem{Preskill:1979zi}
J.~P.~Preskill,
Phys.\ Rev.\ Lett.\  {\bf 43}, 1365 (1979).

\bibitem{Guth:1980zm}
A.~H.~Guth,
Phys.\ Rev.\ D {\bf 23}, 347 (1981).

\bibitem{Kofman:1995fi}
L.~Kofman, A.~D.~Linde and A.~A.~Starobinsky,
Phys.\ Rev.\ Lett.\  {\bf 76}, 1011 (1996)%
\archive{[hep-th/9510119]}.

\bibitem{Khlebnikov:1998sz}
S.~Khlebnikov, L.~Kofman, A.~D.~Linde and I.~Tkachev,
Phys.\ Rev.\ Lett.\  {\bf 81}, 2012 (1998)
[hep-ph/9804425].

\bibitem{Rajantie:2000fd}
A.~Rajantie and E.~J.~Copeland,
Phys.\ Rev.\ Lett.\  {\bf 85}, 916 (2000)%
\archive{[hep-ph/0003025]}.

\bibitem{Kasuya:1998td}
S.~Kasuya and M.~Kawasaki,
Phys.\ Rev.\ D {\bf 58}, 083516 (1998)%
\archive{[hep-ph/9804429]}.

\bibitem{Tkachev:1998dc}
I.~Tkachev, S.~Khlebnikov, L.~Kofman and A.~D.~Linde,
Phys.\ Lett.\ B {\bf 440}, 262 (1998)
[hep-ph/9805209].

\bibitem{Zurek:1985qw}
W.~H.~Zurek,
Nature {\bf 317}, 505 (1985);
Phys.\ Rept.\  {\bf 276}, 177 (1996)%
\archive{[cond-mat/9607135]}. 

\bibitem{Elitzur:im}
S.~Elitzur,
Phys.\ Rev.\ D {\bf 12}, 3978 (1975).

\bibitem{Polyakov:vu}
A.~M.~Polyakov,
Phys.\ Lett.\ B {\bf 72}, 477 (1978).

\bibitem{Hart:1996ac}
A.~Hart, O.~Philipsen, J.~D.~Stack and M.~Teper,
Phys.\ Lett.\ B {\bf 396}, 217 (1997)
[hep-lat/9612021].

\bibitem{Weinberg:uq}
E.~J.~Weinberg,
Phys.\ Lett.\ B {\bf 126}, 441 (1983).

\bibitem{Lee:gt}
K.~M.~Lee and E.~J.~Weinberg,
Nucl.\ Phys.\ B {\bf 246}, 354 (1984).

\bibitem{Digal:1998ak}
S.~Digal, R.~Ray and A.~M.~Srivastava,
Phys.\ Rev.\ Lett.\ {\bf 83} (1999) 5030%
\archive{[hep-ph/9805502]}.


\bibitem{Farakos:1994kx}
K.~Farakos, K.~Kajantie, K.~Rummukainen and M.~E.~Shaposhnikov,
Nucl.\ Phys.\ B {\bf 425}, 67 (1994)
[hep-ph/9404201].

\bibitem{Moore:1996bn}
G.~D.~Moore and N.~Turok,
Phys.\ Rev.\ D {\bf 55}, 6538 (1997)
[hep-ph/9608350].

%

\bibitem{Rajantie:1997pr}
A.~Rajantie,
Nucl.\ Phys.\ B {\bf 501}, 521 (1997)%
\archive{[hep-ph/9702255]}.


\bibitem{Hindmarsh:2001vp}
M.~Hindmarsh and A.~Rajantie,
Phys.\ Rev.\ D {\bf 64}, 065016 (2001)
[hep-ph/0103311].

%

\bibitem{Coleman:jx}
S.~R.~Coleman and E.~Weinberg,
Phys.\ Rev.\ D {\bf 7}, 1888 (1973).

\bibitem{Kajantie:1997tt}
K.~Kajantie, M.~Laine, K.~Rummukainen and M.~E.~Shaposhnikov,
Nucl.\ Phys.\ B {\bf 503}, 357 (1997)
[hep-ph/9704416].


\bibitem{Kronfeld:1987vd}
A.~S.~Kronfeld, G.~Schierholz and U.~J.~Wiese,
Nucl.\ Phys.\ B {\bf 293}, 461 (1987).

\bibitem{Davis:2000kv}
A.~C.~Davis, T.~W.~Kibble, A.~Rajantie and H.~Shanahan,
JHEP {\bf 0011}, 010 (2000)
[hep-lat/0009037].

%

\bibitem{Hindmarsh:2000kd}
M.~Hindmarsh and A.~Rajantie,
Phys.\ Rev.\ Lett.\  {\bf 85}, 4660 (2000)%
\archive{[cond-mat/0007361]}.

\bibitem{Rajantie:2001ps}
A.~Rajantie,
Int.\ J.\ Mod.\ Phys.\ A {\bf 17}, 1 (2002)%
\archive{[hep-ph/0108159]}.

\bibitem{Davis:2001mg}
A.~C.~Davis, A.~Hart, T.~W.~Kibble and A.~Rajantie,
Phys.\ Rev.\ D {\bf 65}, 125008 (2002)
[hep-lat/0110154].

\bibitem{Einhorn:ym}
M.~B.~Einhorn, D.~L.~Stein and D.~Toussaint,
Phys.\ Rev.\ D {\bf 21}, 3295 (1980).



\end{thebibliography}
\end{document}